\documentclass[twocolumn,showpacs,preprintnumbers,amsmath,amssymb]{revtex4}

\usepackage{graphicx}
\usepackage{dcolumn}
\usepackage{bm}
\usepackage{epsfig}
\usepackage{epstopdf}

\begin{document}


\title{Generation of two-photon states with arbitrary degree of entanglement via nonlinear crystal superlattices}

\author{Alfred B. U'Ren$^{1,3,4}$,Reinhard K. Erdmann$^{2,4}$, Manuel de la Cruz-Gutierrez$^{1,4}$, Ian A. Walmsley$^1$}
\affiliation{$^1$Clarendon Laboratory, Oxford University, Parks Road, Oxford, OX1 3PU, England\\
$^2$Sensors Directorate, Air Force Research Laboratory, Rome, NY, USA $\mbox{                                   }$\\
$^3$Centro de Investigaci\'{o}n Cient\'{i}fica y Educaci\'{o}n
Superior de Ensenada (CICESE), Baja California, 22860, Mexico\\
$^4$The Institute of Optics, University of Rochester, Rochester, New
York 14627, USA}

\date{\today}

%
\newcommand{\epsfg}[2]{\centerline{\scalebox{#2}{\epsfbox{#1}}}}

\begin{abstract}
We demonstrate a general method of engineering the joint quantum
state of photon pairs produced in spontaneous parametric
downconversion (PDC). The method makes use of a superlattice
structure of nonlinear and linear materials, in conjunction with a
broadband pump, to manipulate the group delays of the signal and
idler photons relative to the pump pulse, and realizes photon pairs
described by a joint spectral amplitude with arbitrary degree of
entanglement. This method of group delay engineering has the
potential of synthesizing a broad range of states including
factorizable states crucial for quantum networking and states
optimized for Hong-Ou-Mandel interferometry. Experimental results
for the latter case are presented, illustrating the principles of
this approach.
\end{abstract}

\maketitle


Quantum interference between single-photon wavepackets lies at the
heart of novel quantum-enhanced communication and computation
technologies. Such interference depends crucially on the
spatio-temporal mode structure of the photons used. Therefore it is
important to develop techniques that enable the generation of
quantum states with the specific modal character required for
different applications. In particular, entangled photon pairs
provide an important resource for several quantum information
processing (QIP) protocols, and single photons in pure states a
resource for others. While great progress has been made in the
generation of single photons on-demand\cite{michler00},  the process
of parametric downconversion (PDC) constitutes a practical
room-temperature source of non-classical radiation exhibiting
unparalleled flexibility for both sets of applications.

Important classes of entangled two-photon states include those
exhibiting positive frequency
correlations\cite{erdmann00,giovannetti01}, required for certain
metrological applications, and those characterized by vanishing
correlations in all degrees of freedom, crucial for applications
that require the concatenation of several sources, such as linear
optical quantum computation\cite{knill01}. In most PDC sources to
date, however, the photon pair directions and frequencies are
strongly correlated, leading to space-time entanglement in the joint
quantum state. This means that when one photon is detected, the
other is projected into a mixed state, which will not interfere with
other photons\cite{Zhang05,uren05}. Experimental proposals for
generating factorable states have been recently
reported\cite{grice01,uren05,uren03}; these tend to rely on
favorable material properties occurring at specific wavelengths. A
more general method is to engineer the photonic state by coherent
manipulation of the probability amplitudes for generating photons of
particular frequency and/or direction.



The synthesis of two-photon states characterized by a joint
amplitude of arbitrary shape in two-dimensional signal and idler
frequency space $\omega_s$--$\omega_i$ requires the use of a
broadband pump. A narrow bandwidth pump centered at $\omega_p$ with
width $\epsilon$ gives rise to photon pairs whose frequencies are
nearly perfectly anti-correlated such that
$|\omega_s+\omega_i-\omega_p| \le \epsilon/2$. Thus, for example,
broadband states exhibiting vanishing or positive spectral
correlations cannot be constructed within such a one-dimensional
subspace; use of a broadband pump frees us from this restriction.
Work by Klyshko and others\cite{klyshko93,digiuseppe02} has explored
the potential for PDC state engineering of compensating dispersive
effects in nonlinear crystals by appropriately chosen linear media.
Here we build upon this pioneering work, presenting both
theoretically and experimentally a technique which relies on a
``superlattice'' of nonlinear and anisotropic linear segments,
pumped by an ultra-short pulse, which permits the generation of
photon pairs with an arbitrary degree of entanglement via
appropriate group-delay engineering.   When used in waveguided
geometries, where spectral and spatial degrees of freedom are
decoupled\cite{uren04} making large collection apertures possible,
the technique presented here can lead to bright photon pair sources
with customizable modal character.

The key to this method is an extension of the notion of
\textit{group velocity} matching that was introduced for the
generation of spectrally decorrelated two-photon
states~\cite{keller97,grice01}, to the idea of manipulating overall
\textit{group delay} mismatches between the various fields in
structured materials. Let us note that a pulsed pump provides a
timing reference for the PDC photons, not present in the CW case.
Indeed, the ability to arbitrarily select the orientation in
$\omega_s$--$\omega_i$ space of the joint spectral amplitude (thus
permitting an arbitrary degree of entanglement) requires, in the
pulsed case, control over two group delays (pump-signal and
pump-idler) compared to a single group delay (signal-idler) in the
CW case. Therefore, in general, two-dimensional group delay mismatch
control is required, which can be achieved by interspersing
nonlinear superlattice segments with linear segments exhibiting an
appropriate birefringence.

To be concrete, the two-photon state produced by spontaneous PDC for
fixed propagation directions\cite{grice97} is:
\begin{equation}
|\Psi\rangle=A\int\limits_0^\infty\int\limits_0^\infty d\omega_s
d\omega_i f(\omega_s,\omega_i) a_s^\dag (\omega_s) a_i^\dag
(\omega_i)|0\rangle_s|0\rangle_i
\end{equation}

where $A$ is a normalization constant, $a_\mu^\dag(\omega_\mu)$
(with $\mu=s,i$) is the creation operator for the two PDC modes and
$f(\omega_s,\omega_i)$ is the joint spectral amplitude (JSA) given
by the product of the pump envelope function (PEF)
$\alpha(\omega_s+\omega_i)$ and the phasematching function (PMF)
$\phi(\omega_s,\omega_i)$ \cite{grice97}.  The normalized joint
spectral intensity (JSI) $|f(\omega_s,\omega_i)|^2$ represents a
probability density function for PDC emission at specific
frequencies $\omega_s$ and $\omega_i$.

\begin{figure}[h]
\epsfg{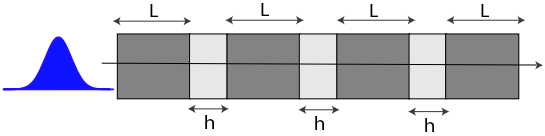}{.4} \caption{Nonlinear crystal superlattice,
pumped by an ultrashort pulse, yielding an arbitrary degree of
entanglement.\label{Fig:XtalSeq}}
\end{figure}

Our source design is based on a periodic assembly of standard
$\chi^{(2)}$ crystals and linear birefringent spacers.  Consider the
PDC source shown in Fig.~\ref{Fig:XtalSeq} consisting of $N$
identical $\chi^{(2)}$ crystals of length $L$ (cut and oriented for
degenerate collinear type-II PDC centered at $\omega_0$) and $N-1$
identical linear $\chi^{(1)}$ spacer crystals of length $h$.  The
PMF for such a superlattice is given by\cite{uren05,klyshko93}:

\begin{equation}\label{E:FullPMF}
\phi_N(\nu_s,\nu_i)\propto
\frac{\mbox{sin}[N\Phi(\nu_s,\nu_i)/2]}{\mbox{sin}[\Phi(\nu_s,\nu_i)/2]}\mbox{sinc}\left[
L \Delta k(\nu_s,\nu_i)/2\right]
\end{equation}
where $\nu_{s,i}=\omega_{s,i}-\omega_0$,
\begin{equation}
\Phi(\nu_s,\nu_i)=L \Delta k(\nu_s,\nu_i)+h \Delta
\kappa(\nu_s,\nu_i),
\end{equation}
and where $\Delta k(\nu_s,\nu_i)$ and  $\Delta \kappa(\nu_s,\nu_i)$
are the phase mismatches in the crystal and spacer segments:
\begin{eqnarray}
\Delta k(\nu_s,\nu_i)&=&k_p(\nu_s+\nu_i)-k_s(\nu_s)-k_i(\nu_i)
\nonumber\\
\Delta \kappa(\nu_s,\nu_i)&=&
\kappa_p(\nu_s+\nu_i)-\kappa_s(\nu_s)-\kappa_i(\nu_i).
\end{eqnarray}

Hence, the total PMF is composed of the product of two separate
functions: one due to each individual crystal segment and one due to
the superlattice. A change of variables to
$\nu_\pm=2^{-\frac{1}{2}}(\nu_s\pm\nu_i)$ and a Taylor expansion of
the crystal and spacer phasemismatches yields:

\begin{equation}\label{E:phi}
\Phi(\nu_+,\nu_-)=h \Delta \kappa^{(0)}+L \Delta
k^{(0)}+\tau_{+}\nu_{+}+\tau_{-}\nu_{-},
\end{equation}

where $\Delta k^{(0)}$ and $\Delta \kappa^{(0)}$ are evaluated at
$\omega_0$ and:
\begin{eqnarray}\label{E:taumtaup}
\tau_{+}&=&2^{-\frac{1}{2}}\left[L(k_s'+k_i'-2k_p')+h(\kappa_s'+\kappa_i'-2\kappa_p')\right]
\nonumber \\
\tau_{-}&=&2^{-\frac{1}{2}}\left[L(k_s'-k_i')+h(\kappa_s'-\kappa_i')\right].
\end{eqnarray}

Here $\kappa_\mu'$ and $k_\mu'(\mu = p, s, i)$ denote the reciprocal
group velocities (RGV) in the spacer and nonlinear material
respectively. We assume that $\Delta k^{(0)}=0$, thus ensuring
phasematching in the crystal segments. The resulting superlattice
JSA can be engineered by noting that whereas the orientation of the
single crystal contribution on the $\nu_s$--$\nu_i$ plane is fixed,
that of the crystal sequence can be adjusted by varying the relative
thicknesses of the crystal and spacer segments\cite{uren05}.  Our
approach is to let the single crystal PMF be much broader than the
spectral structure imposed by the superlattice, so that the latter
determines the resulting two-photon state properties. Under these
circumstances and modelling the PEF $\alpha(\omega_s+\omega_i)$ as a
Gaussian function with width $\sigma$ we obtain:
\begin{equation}\label{E:JSAxtalseq}
f(\nu_{+},\nu_{-})\propto\frac{\mbox{sin}[N\Phi(\nu_{+},\nu_{-})/2]}{\mbox{sin}[\Phi(\nu_{+},\nu_{-})/2]}\mbox{exp}
\left[-2\nu_{+}^2/\sigma^2\right]
\end{equation}
with $\Phi(\nu_{+},\nu_{-})$ given by Eq.~\ref{E:phi}.  Let us note
that rotating by $180^\circ$ subsequent crystal segments, thus
flipping the incidence angle relative to the optic axis, it is
possible to limit the maximum transverse walkoff to its single
cyrstal value (spacer segments exhibit no walkoff for propagation
parallel or orthogonal to the optic axis). Alternatively, a
waveguided geometry evidently suppresses spatial walkoff.
Simultaneous temporal and spatial matching between pump, signal and
idler is indeed a powerful tool for the engineering of two-photon
states.

The JSA given by Eq.~\ref{E:JSAxtalseq} may be adapted to produce
various photonic states. This may be illustrated by an example.
States with symmetric JSAs [i.e.
$f(\omega_s,\omega_i)=f(\omega_i,\omega_s)$] exhibit perfect
visibility in two-photon Hong-Ou-Mandel (HOM) interference, an
essential ingredient in many optical QIP protocols.
Eqns.~\ref{E:JSAxtalseq} and \ref{E:phi} reveal that by imposing the
condition $\tau_{-}=0$, $\nu_{-}$ dependence is suppressed; the
former leads to a symmetric JSA, in particular one exhibiting
frequency anticorrelation (which may be very tight for a long
superlattice) which satisfies the HOM symmetry condition despite the
fact that, individually, each type-II crystal does not. In our
approach, the spectral(temporal) distinguishing information from
individual crystal segments is eliminated by compensating the
crystal signal-idler temporal walkoff $L(k'_s-k'_i)$ by that in the
linear birefringent spacers $h(\kappa'_s-\kappa'_i)$, which ensures
the suppression of the overall signal-idler temporal walkoff term
$\tau_{-}$. Thus, crystal and spacer materials must be chosen with
opposite signed signal-idler walkoffs while segment thicknesses must
be selected appropriately.

\begin{figure*}[ht]
\centering \epsfg{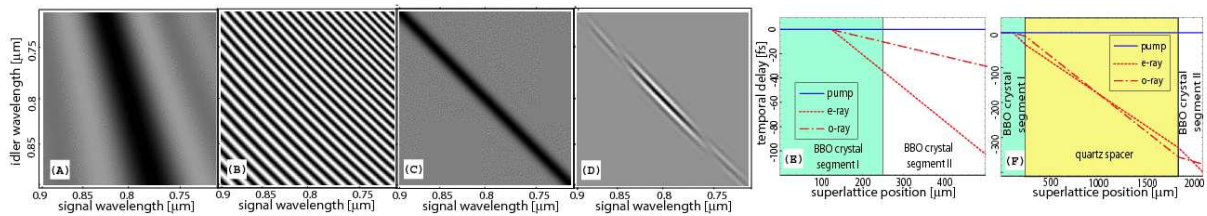}{0.35} \caption{Density
plots of: (A) single crystal segment PMF exhibiting lack of
symmetry, (B) BBO-quartz-BBO superlattice contribution to the PMF,
(C) pump envelope function, (D) JSI symmetrized by the superlattice.
(E)Time (with respect to pump pulse) at which signal/idler generated
halfway through the first crystal traverse each point of two crystal
segments. (F)Same as before for a BBO-quartz-BBO superlattice;
signal-idler walkoff is limited to its single crystal value.
 \label{Fig:XtalSeqHOM}}
\end{figure*}

A JSA of this type can be engineered using BBO crystal segments cut
for degenerate collinear type-II PDC at $800$nm (with $42.4^\circ$
cut angle) and quartz spacers (BBO and quartz fast axes oriented
orthogonally). For BBO segments $k'_e-k'_o=-194$fs/mm, while for
quartz spacers $\kappa'_e-\kappa'_o=31$fs/mm; thus, the spacer to
crystal thickness ratio must be $h/L\approx6.1$ in order that
$\tau_{-}=0$ (the plots in Fig.~\ref{Fig:XtalSeqHOM} assume
$L=250\mu$m and $h=1.53$mm). Fig.~\ref{Fig:XtalSeqHOM}(A) shows the
single crystal PMF, which does not satisfy the HOM symmetry
condition\cite{branning98}. Fig.~\ref{Fig:XtalSeqHOM}(B) shows the
superlattice PMF contribution (left-hand factor in
Eq.~\ref{E:FullPMF}). Fig.~\ref{Fig:XtalSeqHOM}(C) shows the PEF
(centered at $400$nm with $1.95$nm bandwidth).
Fig.~\ref{Fig:XtalSeqHOM}(D) shows the resulting
crystal-spacer-crystal superlattice JSI exhibiting the desired
symmetry; the single crystal spectral bandwidth is sufficiently
large that the superlattice determines the photon pair properties.

The group delay engineering above may be understood by considering
how the quartz spacer limits the signal-idler temporal walkoff.
Fig.~\ref{Fig:XtalSeqHOM}(E) shows for each point along a single
$500\mu$m thick crystal (considered as two $250\mu$m crystals in
sequence) the time, with respect to the pump pulse, at which each of
the signal and idler wavepackets traverses each point of the
crystal, assuming a photon pair created halfway through the first
crystal. Fig.~\ref{Fig:XtalSeqHOM}(F) shows the effect of the quartz
spacer: it limits the maximum signal-idler walkoff to that observed
in a single $250\mu$m crystal (it increases, however, the pump-PDC
walkoff, which does not impact the HOM symmetry condition). Thus,
even though signal-idler temporal walkoff cannot be entirely
suppressed, it can be limited in a superlattice to an arbitrarily
small value.

\begin{figure}[t]
\centering \epsfg{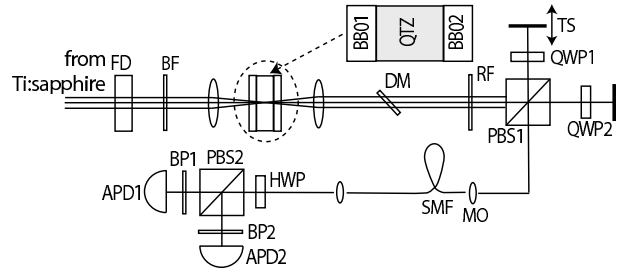}{0.55} \caption{BBO1/BBO2: $250\mu$m
BBO crystals, QTZ: $1.53$mm quartz spacer,  PBS1/PBS2: polarizing
beam splitters,  QWP1/QWP2: quarter waveplates, TS: translatable
mirror, MO: $16\times$ microscope objetive, SMF: single mode fiber,
HWP: half wave plate, BP1/BP2: bandpass filters of $40$nm width,
APD1/APD2: avalanche photo diodes. \label{Fig:setup}}
\end{figure}

In order to illustrate the power of this concept, we implemented the
simple superlattice described above. The experimental configuration
used  [see Fig.~\ref{Fig:setup}] relies on a collinear type II HOM
interferometer (HOMI). A train of ultrashort pulses centered at
$405$nm ($2$nm FWHM bandwidth, $82$MHz repetition rate and $150$mW
power), obtained by frequency doubling a mode-locked Ti:Sapphire
beam, was weakly focused onto a collinear type II PDC source.  The
latter is based on 250$\mu$m length BBO crystal segments, and could
be configured as single-crystal, double-crystal or as a
crystal-quartz-crystal superlattice. A polarizing beamsplitter
(PBS1) established a Michelson interferometer, each polarization in
one arm. Quarter wave plates (QWP1/QWP2) reverse the polarization,
so that both photons exit PBS1 from its second port. One of the arm
lengths was adjustable via a motorized translation stage (TS). The
PDC exiting the interferometer was coupled into a single mode fiber
(SMF) which ensured a high degree of signal-idler spatial
mode-matching, prior to which a dichroic beamsplitter (blue
reflecting/red transmitting) followed by a red-colored glass filter,
separated the PDC light from the blue pump.  A half-wave plate (HWP)
oriented the orthogonal signal/idler modes at $45^{\circ}$ with
respect to the axes of a second polarizing beam splitter (PBS2)
which formed the basis for a collinear HOMI. Each output mode from
the PBS was focused on the active area of a silicon photon-counting
module; wide-bandpass filters (wider than the PDC bandwidth) were
used for background suppression. Data was taken by monitoring
coincidence counts as a function of the translation stage (TS)
position.

\begin{figure}[t]
\centering \epsfg{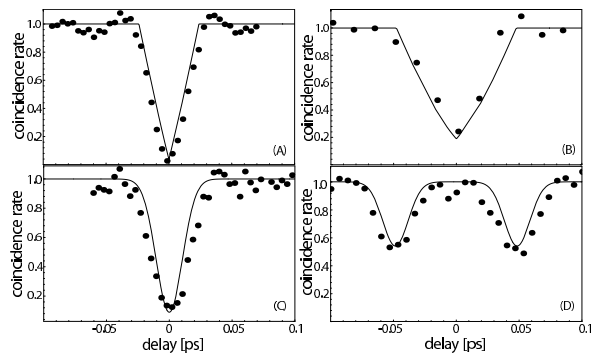}{0.54} \caption{ HOM
interference curves, normalized to a unit background to suppress
source brightness variation resulting from re-alignment for each
experiment. (A) single crystal segment. (B) Two crystal segments in
sequence, exhibiting reduced visibility and broadening. (C)
BBO-quartz-BBO superlattice exhibiting restoration of visibility.
(D) BBO-quartz-BBO with quartz rotated by $90^\circ$; visibility no
longer restored.\label{Fig:DataFig2}}
\end{figure}

The data in Fig.~\ref{Fig:DataFig2} shows coincidence count rate
data, normalized to a unit background, as a function of the
signal-idler relative delay. Solid curves represent theoretical
curves obtained from a calculation similar to that in
Ref.~\cite{grice97}. Fig.~\ref{Fig:DataFig2}(A) shows the
interference curve resulting from a single crystal segment.
Fig.~\ref{Fig:DataFig2}(B) shows a broadened dip with reduced
visibility, due to the increased signal-idler temporal walkoff
[quantified by $\tau_{-}$, Eq.~\ref{E:taumtaup}], resulting from two
crystal segments in sequence. Fig.~\ref{Fig:DataFig2}(C) shows the
result of placing a $1.6$mm length quartz spacer between the
crystals [see Fig.~\ref{Fig:setup}]. Note that the visibility and
dip width are restored to their single crystal values, thus
demonstrating the effect of limiting the maximum walkoff to that
experienced by a single crystal. Fig.~\ref{Fig:DataFig2}(D) shows
the effect of rotating the quartz spacer by $90^\circ$, thus
reversing the extraordinary and ordinary character in the spacer. In
this latter case, the superlattice contribution is rotated away from
the diagonal axis on the $\omega_s$--$\omega_i$ plane, thus
destroying fulfilment of the HOM symmetry condition; the
interference curve now contains two dips, each associated with one
of the crystals.  Let us note that our conclusions depend on an
analysis of HOM interference features and are independent of source
brightness; however, a future refinement is the careful selection of
the spacer thickness in order to rule out thickness-dependent count
rate oscillations \cite{herzog04}.

So far we have considered the effect of signal-idler temporal
walkoff. In general the pump experiences a different group velocity
from the signal and idler photons, so that there is a corresponding
temporal walkoff between the pump and the PDC photon pairs.  It is
evident that the condition $\tau_{+}=0$ (see Eq.~\ref{E:taumtaup})
closely resembles the group velocity matching (GVM)
condition\cite{keller97,grice01} by which the pump RGV is matched to
the mean RGV of the signal and idler wavepackets. In a superlattice,
however, it is the total group delay mismatch which must vanish.
This generalized condition can lead to spectrally factorable states,
or to states exhibiting positive spectral correlations. For example,
$\tau_{+}=0$ can be satisfied using type-II BBO crystal segments and
calcite spacers\cite{uren05}. Note that the GVM condition fulfilled
by the superlattice depends crucially on $L$ and $h$.  The thickness
error tolerance becomes more restrictive for longer assemblies (i.e.
for higher $N$); for example for a superlattice of 10 BBO crystals
and 9 calcite spacers designed to yield a positive correlation
state, the tolerance is about 2\%. In addition, the technique
presented here can be adapted to produce states exhibiting a very
large Schmidt number (and thus a high degree of space-time
entanglement)\cite{zhang05b} which constitute extreme examples of
non-classical correlations\cite{law04}.




In conclusion, we have shown that a periodic array of nonlinear
crystals and linear birefringent spacers, pumped by an ultrashort
pump, can result in PDC photon pairs exhibiting an engineered JSA
function. We have illustrated this concept by demonstrating a source
capable of high-visibility HOM interference, in a type-II pulsed
pump configuration, using using materials that do not generate such
states naturally. Our scheme has wider applicability; it can
generate a broad class of two photon states including those with
factorable and positively correlated JSAs, as well as those with a
very large Schmidt number. It is conceivable that appropriate
superlattices could be implemented in monolithic nonlinear optical
structures where locally-modified dispersion characteristics are
obtained by ion-exchange, doping or other mechanisms. This would
open up a powerful new source engineering technique, akin to quasi
phasematching but in the group velocity, rather than phase velocity,
domain.

\begin{acknowledgements}
We acknowledge useful conversations with M. Fejer and M.G. Raymer.
This work was supported by ARDA grant P-43513-PH-QCO-02107-1. AU
acknowledges support from the Center for Quantum Information, funded
by ARO administered MURI grant DAAG-19-99-1-0125.
\end{acknowledgements}


\end{document}